\documentclass[11pt]{amsart}
\usepackage{amsmath,amssymb} 
\usepackage{latexsym}
\usepackage{diagrams} 
\newtheorem{thm}{Theorem}[section]
\newtheorem{cor}[thm]{Corollary}
\newtheorem{conj}[thm]{Conjecture}
\newtheorem{lem}[thm]{Lemma}
\newtheorem{prop}[thm]{Proposition}
\theoremstyle{definition}
\newtheorem{deff}{Definition}[section]
\theoremstyle{remark}
\newtheorem{rem}{Remark}[section]
\numberwithin{equation}{section}
\newcommand{\thmref}[1]{Theorem~\ref{#1}}
\newcommand{\secref}[1]{\S\ref{#1}}
\newcommand{\lemref}[1]{Lemma~\ref{#1}}
\newcommand{\propref}[1]{Proposition~\ref{#1}}

\newcommand{\remref}[1]{Remark~\ref{#1}}

\newcommand{\nc}{\newcommand}
\nc{\bib}{\bibitem}
\nc{\on}{\operatorname}
\nc{\res}{\operatornamewithlimits{Res}^\cdot}
\nc{\resold}{\operatornamewithlimits{Res}}
\nc{\tod}{\on{Tod}}
\nc{\ahat}{\on{Ahat}}
\nc{\td}{\on{td}}
\nc{\aht}{\on{ahat}}
\nc{\ul}{\underline}
\nc{\arr}{\rightarrow}
\nc{\al}{\alpha}
\nc{\C}{{\mathbb C}}
\nc{\Cn}{{\mathbb C}^n}
\nc{\Z}{{\mathbb Z}}
\nc{\R}{{\mathbb R}}
\nc{\bb}{{\bar B}}
\nc{\A}{\mathfrak{A}}
\nc{\I}{\mathfrak{I}}
\nc{\eps}{\varepsilon}
\nc{\W}{\mathfrak{W}}
\nc{\M}{\mathfrak{M}}
\nc{\ml}{\ll}
\nc{\mgg}{\M^G_g}
\nc{\vs}{V^*}
\nc{\Sym}{\on{Sym}}
\nc{\kh}{\eta}

\nc{\Hom}{\on{H}}
\nc{\tensor}{\otimes}
\nc{\spf}{$\operatorname{Spin}(5)$~}
\nc{\suk}{$\operatorname{SU}(2)$~}
\nc{\suh}{$\operatorname{SU}(3)$~}
\nc{\sun}{$\operatorname{SU}(n)$~}
\nc{\ssum}{\sum}
\nc{\im}{\on{im}}
\nc{\rat}{R_\A}
\nc{\merom}{M_\A}  
\nc{\Mu}{\Gamma_\A}
\nc{\irm}{\on{I}} 
\nc{\hirm}{\widehat{\on{IRM}}}
\nc{\ir}{\mathrm{IR}}
\nc{\hir}{\widehat{\on{IR}}}
\nc{\os}{b}
\nc{\fl}{\on{fl}}
\nc{\FLA}{\mathrm{FL}_\A}
\nc{\bcb}{\on{BCB}}
\nc{\rell}{\on{Rel}}
\nc{\rel}{\on{rel}}
\nc{\End}{\on{End}}
\nc{\id}{\mathrm{id}}
\nc{\ach}{\A^*}  
\nc{\starr}{$*$}
\nc{\Or}{\A_\mathrm{ind}} 
\nc{\Rf}{\Z \Or} 
\nc{\OS}{\mathbf{OS}}
\nc{\DOS}{\widehat{\mathbf{OS}}}
\nc{\hm}{\hat M}
\nc{\ho}{\hat O}
\nc{\ha}{\hat{\mathfrak A}}
\nc{\hos}{\hat \mu}
\nc{\todd}{\mathrm{Todd}}
\nc{\vex}{\vec\bx}
\nc{\vol}{\on{vol}}
\nc{\cube}{\Box}
\nc{\oma}{\Omega_{\A}}
\nc{\less}{\backslash}
\nc{\ba}{\bigwedge \Z\A}
\nc{\mer}{\on{Mer}_\A}
\nc{\Cone}{\on{Cone}}
\nc{\cla}{\mathbb{E}(\Lambda)}
\nc{\bx}{\mathbf{x}}
\nc{\rt}{\hat R_\A}
\nc{\ttt}{\tilde t}
\nc{\dn}{\mathfrak{D}_n}
\nc{\bbb}{BB^n}
\nc{\bara}{\overline{|\A|}}
\nc{\abez}{\A\backslash H}
\nc{\diag}{\mathrm{diag}}
\nc{\diagd}{\widehat{\mathrm{diag}}}
\nc{\bbeta}{\mathbf{a}}
\nc{\mudel}{\hat{\mu}_\Delta}
\nc{\qdel}{\hat{q}_\Delta}
\nc{\hatir}{\widehat{\ir}_\Delta}
\nc{\bz}{\vec{\mathbf{z}}}
\nc{\ires}{\operatornamewithlimits{IRes}}

\begin{document}

\title{Iterated Residues and Multiple Bernoulli Polynomials}
\author{Andr\'as Szenes}
\thanks{The research was supported by the Alfred Sloan Foundation and the
Mittag-Leffler Institute}
\address{Massachusetts Institute of Technology, Department of Mathematics}
\email{szenes@math.mit.edu}

\maketitle

\section{Introduction} \label{intro}
In this paper we consider the following
 problem: let $\A$ be a central hyperplane
 arrangement over $\Z$; equivalently, let $V$ be an $n$-dimensional real
 vector space and $\Gamma\subset V$ a lattice of rank $n$.  We will always
 assume that the $\cap\A=\{0\}$ and that the lines dual to the hyperplanes
 in $\A$ span $V^*$. Denote by $\Lambda$ the lattice in $\vs$ dual to
 $\Gamma$ over $\Z$.  The integrality condition says that each hyperplane
 $H$ in the arrangement is the zero set of some linear form $x_H\in
 \Lambda$. We will call a real hyperplane arrangement (HPA) with this extra
 structure {\em integral} (IHPA).  Of course, up to isomorphism we always
 have $\Z^n=\Lambda\subset V=\R^n$.

Some more {\em notation}: Let $U=V\backslash \cup\A$ be the complement
of the arrangement and $\Mu=U\cap\Gamma$.  To avoid factors of $2\pi
\sqrt{-1} $ in the formulas, introduce $\ul j=j/(2\pi \sqrt{-1})$. We
call a function on $V$ $\A$-{\em rational} if it is rational and its
poles are contained in $\cup\A$; similarly an $\A$-{\em meromorphic}
function is a meromorphic function defined in a neighborhood of $0\in
V$ with poles contained in $\cup\A$. Denote these classes of functions
by $\rat$ and $\merom$.

For any $\A$-rational function $f$, we are interested in calculating the
following sums:
\begin{equation}
B_{f,\A} = \ssum_{\ul j\in \Mu} f(j).
\end{equation}
An example of such a sum is
\[ S(a,b,c) = \sum \frac 1{m^an^b(m+n)^c},\quad m,n\in\Z;\;m,n,m+n\neq0,
\]
where $a,b,c$ are positive integers.

More generally we will consider the Fourier series: 
\begin{equation} \label{bdef}
B_{f,\A}(t) = \ssum_{\ul j\in\Mu} \exp\langle t,j\rangle f(j),
\end{equation}
where $t\in \vs$.

We will omit $\A$ from the notation whenever it does not cause confusion.
The infinite sum $B_{f,\A}(t)$ makes sense as a distribution. For now, we
assume that these sums converge absolutely, thus $B_{f,\A}(t)$ is a
continuous function of $t$, although eventually we will consider all
functions $f\in\rat$.

A few basic properties: clearly, $B_f(t)$ is
$\Lambda$-periodic in $t$ and $B_f(0)=B_f$.  We will show that that $\vs$
has a chamber decomposition such that $B_f(t)$ restricts to a {\em polynomial}
inside each chamber.  Because of the periodicity, this decomposition comes
from one of $T=\vs/\Lambda$. These polynomials are natural generalizations
of the classical Bernoulli polynomials, which correspond to the
one-dimensional case (see \secref{onedimpol}). When $f$ has rational
coefficients, these polynomials will have rational coefficients as
well. Following Zagier \cite{zagier} we named them {\em multiple Bernoulli
polynomials} (MBP).

Now we describe the shape of the result in general terms.  The main
goal is to give residue formulas, a certain type of generating series,
for these MBPs. The singular locus of the MBPs are given by the
$\Lambda$-translates of the dual arrangement $\ach$ in $\vs$ formed by
the one-dimensional faces of $\A$. Thus for every $f\in\rat$ and every
chamber $\Delta$ of $\Lambda+\ach$ we have to find a polynomial
$B^\Delta_f(t)$ such that $B^\Delta_f(t)=B_f(t)$, whenever $t\in
\Delta$.

The type of residues which appear naturally in our approach are the {\em
iterated residues} defined in \secref{itres}. One of the features of our
formula is that it is independent of $f$, i.e.  for any integral
arrangement $\A$ and chamber $\Delta$, we construct a local residue form
$\omega_\Delta(\A)$ such that whenever $t\in\Delta$ we have
$$B^\Delta_f(t) = \langle \omega_\Delta(\A), f \exp(t)\rangle.$$
The precise definition of the pairing $\langle,\rangle$ is given in
\eqref{pairing}. One can think of the collection
$\omega=\{\omega_\Delta(\A)\}$ as of a fundamental class of the IHPA $\A$.

The motivation for considering such sums came from the work of Witten on
the intersection ring and volume of the moduli spaces of $G$-bundles on
Riemann surfaces \cite{witten,kef}, where $G$ is a compact connected Lie
group; the arrangement is given by the coroot hyperplanes in the dual of
the Cartan subalgebra of $G$; the lattice $\Gamma$ is the weight lattice of
the group. Some details are given in \secref{applic}.

The existence of such $\omega$ in the Lie group case mentioned above was
conjectured in \cite[Conjecture 4.2]{me} in a weaker form. We will
discuss the relevance of this statement to the topology of the moduli
spaces in a separate paper.  The sketch of the proof of this conjecture was
given in \cite{me} for $G=SU(3)$, with the claim that the case of \sun is
analogous. This result was later used and extended in the works of Jeffrey
and Kirwan \cite{jef} (see also \cite{mori}). The Verlinde-type deformation
of these formulas (cf. \cite{me} for a detailed explanation) will be given
in a future publication.

Naturally, besides the above, an explicit construction of $\omega$ is also
useful since it gives a simple, efficient  calculus for the numbers
$B_f$. A different approach to the calculation of these numbers was
given earlier by Zagier \cite{zagier}. While the formulas in \cite{zagier}
are not as compact as ours, there the multiplicative structure of these
numbers and even certain modular deformations of them are considered.

The contents of the paper: In \secref{onedimpol} we describe the classical
theory in terms of our general setup, in \secref{itres} we introduce our
version of iterated residues -- a very special case of the standard notions
such as the Grothendieck residue and Parshin's residue and show how it
relates to the theory of hyperplane arrangements. We give a new
interpretation of the {\em broken circuit bases} along the way. In
\secref{integral} we prove the main theorem while in \secref{applic} we
give some applications and calculations; we work out the most important
example of the braid arrangements.

{\bf Acknowledgments}. The first version of this paper (hep-th/9707114) was
completed during my stay at the Mittag-Leffler Institute in the Spring of
1997. I am most grateful to Lev Rozansky and Kefeng Liu whose helpful
comments and interest made this work possible. I would like to thank Victor
Batyrev, Hiroaki Terao, Alexandr Varchenko, Michel Vergne and Andrei
Zelevinski for their suggestions and encouragement. Eva Maria Feichtner
provided me with some important references.

\section{Classical Bernoulli polynomials} \label{onedimpol}

The classical Bernoulli polynomials can be
defined using the setup of the introduction as follows: let $V=\R$, $\Gamma=\Z$;
there is one ``hyperplane'' -- $\{0\}$, $\Mu=\Z\backslash\{0\}$. 

By abuse of notation (we replace $x^{-k}$ by $k$ in the subscript of $B$)
we have 
\begin{equation} \label{defin}
B_k(t) = \ssum_{\ul n\in\Mu} \frac{\exp(nt)}{n^k}.
\end{equation}
This function is defined by a Fourier series, therefore,  it is
periodic with period 1. Properties:
\begin{enumerate}
\item From the Fourier series it is clear that $B_k(t)$ is $k-2$
times differentiable.
\item $\frac{d}{dt} B_k(t)=B_{k-1}(t)$.
\item $B_0(t) = \delta_\Z(t)-1$, where  $\delta_\Z(t)$ is the
sum of delta-functions at all integers.
\item $\int_0^1 B_k(t)\, dt=0$, by inspecting the Fourier series. 
\item It follows from Properties 1--4 that $B_k$ restricts to a
polynomial of degree exactly $k$ on each interval $(m,m+1)$, which will be
denoted by $B_k^m$.   These polynomials are
defined recursively and have rational coefficients.
\end{enumerate}
The polynomials $B_k^0(t)$
are the Bernoulli polynomials up to a sign and a factor of $k!$.

\subsection{Residue formulas}
The main goal of this paper is to obtain residue formulas for $B_k(t)$
and its generalizations. The basic technical tool is
\begin{lem} \label{onedim} 
Let $f$ be a rational function of degree $\leq-2$ on  $\C$ and let $P$ 
be the set of its poles. Then for each $t$, $0\leq t <1$,
\begin{equation}
\ssum_{\ul n\in\Z\backslash P} \exp(nt) f(n) = \sum_{p\in P} \resold_{x=p}
\exp(tx)\td(x) f(x),
\end{equation}
where
$$\td(x) = \frac{dx}{1-\exp(x)}.$$
\end{lem}
\begin{proof}
Apply the Residue Theorem to a circle of radius $\pi L$ for odd integer
$L$ and let $L\arr \infty$. 
\end{proof}

This immediately gives us a residue formula for the Bernoulli polynomials:
denoting by $\{t\}$  the fractional part of $t$
\begin{equation} \label{kg2}
B_k^0(t) = \resold_{x=0} \exp(\{t\}x)\td(x) x^{-k},
\end{equation}
at least for $k\geq 2$. 

Note that for any $k$ the series $B_k(t)$ can be interpreted as a
distribution. Then  by differentiating $B_k(t)$ considered as a
distribution, we can extend \eqref{kg2}:
\begin{lem} \label{distribution}
For arbitrary integer $k$ the distribution $B_k(t)$ is a polynomial
function in each interval $(m,m+1)$, $m\in\Z$. For a non-integral $t$ we
have
$$B_k^m(t) = \resold_{x=0} \omega^m x^{-k},$$
where $\omega^m=\exp(-mx)\td(x)$. 
\end{lem}

Note that $B_1(t)$ is a discontinuous function and $B_1^0(t) = 1/2-t$, while
$B_0(t)$ is a distribution and $B_0^m(t)=-1$ for all $m$. For $k<0$ always
$B_k^m=0$.

\section{Iterated Residues and Hyperplane Arrangements}  \label{itres}

\subsection{Residue Forms}  In this section, let $\A$ be a complex
HPA in an $n$-dimensional complex vector space $V_\C$ with complement
$U_\C$.  Consider a linearly independent ordered n-tuple
$S=(H_1,H_2,\dots,H_n)$ of hyperplanes in $\A$. We define a linear
operation on $\A$-meromorphic functions $\ires_S:\merom\arr\C$ as
follows. First note that there is a well defined notion of ``the constant
term at 0'' of a meromorphic function $f$ on a complex line (1-dimensional
complex vector space), which we will denote by $\res_{z=0}g$. This is
simply the degree 0 coefficient in the Laurent expansion of $f$. Take an
element $g\in\rat$ and consider the family of lines parallel to
$\cap_{i=2}^{n} H_i$.  Each of these lines is a complex line with 0 on
$H_1$ and the restriction of $g$ onto them has a well-defined constant
term. Thus we obtain a new function $\res_{H_1} g$ on $H_1$. This procedure
can be repeated with $H_2$, replacing $V_\C$ by $H_1$, $g$ by $\res_{H_1}
g$ and $H_i$ by by $H_1\cap H_i$. Iterating this $n$ times, we arrive at a
number
$$\ires_S g = \res_{H_n}\dots\res_{H_2}\res_{H_1} g.$$
We call this operation an iterated residue. A few important points:
\begin{rem} \label{restrict} 
\begin{enumerate}
\item Usually taking a residue is applied to differential forms not
instead of $\resold$ to mark the difference, but it may have been more
proper to call this operation ``constant term'' rather than residue.
\item  The operation $\ires_S$ is a degree 0 operation with respect to
the $\C^*$ action on $V_\C$.
\item A more practical method of computation of iterated residues is given
in \secref{applic}.
\item  The operation $\ires_Sg$ depends on the order of the hyperplanes!
\item The notation $\res_{H_1} g$ above is inconsistent since this
operation actually depends on the other $H_i$-s as well. There is one
case, however, when this is not so: when $g$ is regular along $H$,
then $\res_{H_1} g = g|H_1$, the restriction of $g$ onto
$H_1$.
\end{enumerate}
\end{rem}

Now we explain how this operation is related to the standard calculus of
homology and cohomology of the complement of HPAs \cite{orlik,varsch}.  Let
$\Or^k$ be the set of ordered linearly independent $k$-tuples of elements
of $\A$ and let $\Rf^k$ be the free $\Z$-module generated by $\Or^k$. Any
map $m:\Or^k\arr A$ to an Abelian group $A$ extends to a map $m:\Rf^k\arr
A$ denoted by the same letter. Denote by $\oma^*$ the graded algebra of
local meromorphic differential forms on $V_\C$ with poles along $\cup\A$.

Consider an ordered $k$-tuple of hyperplanes $S=(H_1,\dots,H_k)\in \Or^k$,
and assume that $H_i$ is represented by a form $x_i$ and let
$\alpha_i=d\log x_i$, $i=1,\dots,k$. Denote by $\mu(S)$ the differential
form $\alpha_1\wedge\dots\wedge\alpha_k$.  Then $\mu$ is a map from a
$\Z\Or^k$ to $\oma^k$, and it clearly induces a map $\mu^\wedge$ from the
$k$th exterior product of $\Z\A$ to differential $k$-forms. All these forms
are integral and closed and the fundamental statement in the theory is that $\mu$
factors through a surjective map $b$ to $H^k(U_\C,\Z)$.

The story is summarized in the commutative diagram below. The maps
$b^\wedge, \mu^\wedge$ and $q$ are  {\em algebra homomorphisms}.

\newarrow{Onto} ----{>>}
\newarrow{Dashto} {}{dash}{}{dash}>
\newarrow{Into} C--->
$$ \begin{diagram}
        &                               &       \oma^*          &                               &               \\
        &\ruTo(2,4)^\mu         &       \uInto^q        &\luTo(2,4)^{\mu^\wedge}        &               \\
        &                               &        H^*(U_\C,\Z)   &                               &               \\
        &\ruTo^b                        &                       &\luOnto^{b^\wedge}             &               \\
\Rf^*   &                               & \rTo^\wedge   &                               &\wedge^*\Z\A   \\
\end{diagram}
$$
To simplify the notation we will not put an index on the maps marking the
degree.
\begin{rem}
The subalgebra of differential forms generated by logarithmic forms $\OS^* =
\mu(\Rf^*)$ is naturally isomorphic to $H^*(U_\C,\Z)$.  This algebra is
called the {\em Orlik-Solomon algebra} of $\A$ and it can be defined
combinatorially in a more general framework. We cannot do justice
to the subject here and refer the reader to the monograph \cite{orlik} and
the original works of Arnold \cite{arnold}, Brieskorn \cite{brieskorn},
Orlik and Solomon \cite{os} and  Bj\"orner \cite{bjorner}.
\end{rem}

There is an interesting picture related to the homology of the complement
as well. Denote by $\FLA^k$ the set of partial $k$-flags (sequences of
subspaces of codimensions $1,\dots, k$) of $\A$ and again
consider an element $S\in \Or^k$ as above. Then we can associate to $S$ an
element $\fl(S) = (H_1,H_1\cap H_2,\dots,\cap_{i=1}^k H_i)$

One can also associate to $S$ a $k$-cycle represented by a $k$-torus as
follows (cf p. 159 in \cite{varsch}):
fix a set of positive $\eps_i$, $i=1,\dots,n$ such that
$0<\eps_1\ll\eps_2\ll\dots\ll\eps_n$ and and let
$(H_1,\dots,H_k,H_{k+1},\dots,H_n)$ be a completion of $S$ to an
independent $n$-tuple. Then the homotopy class of the torus
\[ \{|x_i| = \eps_i, \, i = 1,\dots, k,\; x_i=\eps_i, \, i=k+1,\dots,n\}
\]
in $U_\C$ is well-defined, and its orientation is fixed by the complex
structure.   The homology of this torus, $Z(S)$,  only depends on $\fl(S)$.
\begin{equation} \label{homology}
\begin{diagram}
        &               &        H_k(U_{\C},\Z) &                       &       \\
        &\ruTo^Z        &                       &\luTo^{Z^{\fl}}        &       \\
\Rf^k   &               &       \rTo^\fl        &                       & \FLA^k\\
\end{diagram}
\end{equation}

One can give explicit generators for the kernels of the maps $b^\wedge$ and
$Z^{\fl}$ (cf. \propref{kermuw} and \cite{varsch})  

By tensoring the maps in the above diagrams we obtain
\begin{equation*} 
\begin{diagram}
        &                       &\oma^k\tensor H_k(U_{\C},\Z)   &                               &       \\
        &\ruTo^{\ir=\mu\tensor Z}       &                               &\luTo^{q\tensor \id=q} &       \\
\Rf^k   &                       &\rTo^{b\tensor Z}              &       & H^k(U_{\C},\Z)\tensor H_k(U_{\C},\Z)\\
\end{diagram}
\end{equation*}

We introduce the notation $\ir$ for the {\em iterated residue map}
$\mu\tensor Z$, and use the same symbol $q$ for the map $q\tensor \id$.


The connection of $\ir$ with iterated residues can be seen as follows:
let $k=n$ and define a pairing between $H_n(U_\C,\Z)\tensor \oma^n$ and $\mer$ by
\begin{equation} \label{pairing}
\langle C\tensor\eta,g\rangle = \int_C g\eta,
\end{equation}
where $C\in H_n(U_\C,\Z)$ and $\eta\in \oma^n$. Note that it is essential
that we restrict to $k=n$ here, because only in this case is $g\eta$
necessarily a closed form.

 Then we have
\begin{prop} \label{simply} 
Let $S\in \Or^n$ and $g\in \mer$. Then
$$ \ires_S g=\langle \ir(S),g\rangle.$$
\end{prop}
{\em Proof}: The proof easily follows from Cauchy's theorem. The residue
integrals which appear in calculating the pairing give exactly the same
result as the procedure in the definition of the iterated residues.

\begin{rem} \label{remark}
The last diagram suggests an interesting application of this 
formalism. Since $H^k(U_{\C},\Z)$ and $H_k(U_{\C},\Z)$ are naturally dual
to each other, there is a canonical diagonal element $\diag^k\in
H^k(U_{\C},\Z)\tensor H_k(U_{\C},\Z)$. Thus there is an invariantly defined
notion of the ``constant term'' of any $f\in\mer$ defined by the formula
\[\langle q(\diag^n),f\rangle.\]
This will be more exciting if we show that this is an iterated residue,
i.e. that $\diag^n\in\im(b\tensor Z)$. This is indeed the case as we will
prove in the next paragraph (cf. \propref{id}).
\end{rem}
We make a much stronger conjecture:
\begin{conj}
The map $b\tensor Z:\Rf^n\longrightarrow  H^n(U_{\C},\Z)\tensor H_n(U_{\C},\Z)$
is surjective.
\end{conj}

\subsection{Broken circuit bases}
Here we recall some of the basics of the theory of HPAs and broken circuit
bases of Orlik-Solomon algebras (cf.\cite{orlik}). 
We show how ``non-commutative'' broken circuit bases fit naturally into the
picture of the previous paragraph.

A {\em circuit} $C$ is a minimally dependent subset of $\A$, i.e. $C$ is
dependent but $C\less\{H\}$ is independent for any $H\in C$. The kernel of
the map $\mu^\wedge:\ba^*\arr\oma^*$ can be described using circuits as
follows: Assume that the elements of a circuit $C$ are linearly ordered and
denote by $C[i]\in\Or^{|C|-1}$ the ordered set obtained by omitting the
$i$th element of $C$.
\begin{prop}[\cite{orlik}] \label{circuit}
Every ordered circuit $C$ gives rise to an element $\rel(C)\in \ker b$
given by 
\begin{equation} \label{relation}
\rel(C) = \sum_{i=1}^{|C|} (-1)^i C[i].
\end{equation}
\end{prop}
\begin{proof}
We give a proof of this statement here, as it will be needed later.

Recall that 
$b$ takes the $k$-tuple $(H_1,\dots,H_k)$ to the form
$\al_1\wedge\dots\wedge\al_k$. Since $\al_i=d\log(x_i)$ depends on
$x_i$ only up to a scalar multiple, and using that $C$ is a circuit, we can
assume that the relation among the elements of $C$ is given by $\sum x_i
=0$. Then for $1\leq s,t\leq k$ we have 
$\wedge_{i\neq s} \,dx_i = (-1)^{s-t} \wedge_{i\neq t} \,dx_i,$
and thus $\mu(\rel(C))$ reduces to
$$ \left(\sum_{i=1}^k\prod_{j\neq i} x_j^{-1}\right) \, \wedge_{i=1}^{k-1} dx_i.$$
This clearly vanishes since the coefficient reduces to $\sum x_i$ after
multiplication by $\prod x_i$.
\end{proof}

\begin{prop}[\cite{orlik}] \label{kermuw}
The  kernel of $\mu^\wedge$ is given by the ideal in $\ba^*$ generated by 
the elements $\{\wedge(\rel(C))|\; C \mathrm{is\; a\; circuit}\}$.
\end{prop}
See \cite{orlik} for the proof.

There is a way to construct a basis of the Orlik-Solomon algebra by
``breaking'' the relations given by the circuits. Denote by $|\A|$ the
number of elements in $\A$, and by $\bara$ the set of integers from 1 to
$|\A|$.  Then one can represent an ordering $\sigma$ of the elements of
$\A$ by a bijection $\sigma:\A\arr\bara$. Below, we define special subsets
$\bcb_\sigma\subset\Or^k$ for every ordering $\sigma$ and every $k$.
To simplify the notation, we will regard elements $S\in\Or^k$ as functions
$S:\{1,\dots,k\}\arr \A$, but, when it does not cause confusion, we will
use the notation $H\in S$ to say that $H$ is one of the hyperplanes in $S$.
\begin{deff} $\bcb^k_\sigma$ consists of those elements of $\Or^k$ ordered
according to $\sigma$ which do not contain ``broken circuits'':
\begin{multline}
\bcb_\sigma^k = \{S\in\Or^k|\,\sigma(S(i))<\sigma((S(i+1)), \quad
1\leq i<k\quad\text{and} \\
\text{for any}\, H\notin S \quad \{H\}\cup\{G\in S|\,
\sigma(G)>\sigma(H)\}\;\text{is independent}\}
\end{multline} 
\end{deff}

Note that our definition is somewhat differs from the standard one in that
we we retain the ordering among the elements of the BCB.

Using BCBs one can construct bases of the Orlik-Solomon algebra: 
\begin{prop}[\cite{orlik}Theorem 3.43]  For any ordering
$\sigma$, the set $\mu(\bcb^k_\sigma)$ is a $\Z$-basis of $\OS^k$ and,
equivalently, $b(\bcb^k_\sigma)$ is a basis of $H^k(U_\C,\Z)$.
\end{prop}

Our ``non-commutative'' version of broken circuit bases has a stronger
property. This is the statement mentioned in Remark \ref{remark}.
\begin{prop} \label{id}
For every ordering $\sigma$ introduce the element 
$$D_\sigma^k=\sum_{S\in\bcb_\sigma} S \in \Rf^k, \quad (1\leq k \leq n).$$
Then 
\[
(b\tensor Z)(D^k_\sigma)=\diag^k.
\]
\end{prop}
\begin{proof}
Let $S,Q\in \Or^k$, and for a permutation $\tau\in S_k$ denote by
$S^\tau$ the n-tuple in $\Or^k$ with the same elements as $S$, but
permuted by $\tau$. Then \cite[\S4]{varsch}
$$\langle b(S),Z(Q)\rangle = 
\begin{cases} 0, & \text{if } \fl(S^\tau)\neq \fl(Q)
\text{ for any } \tau\in S_k\\
(-1)^{\on{parity}(\tau)}, & \text{if } \fl(S^\tau)=\fl(Q), \end{cases}
$$
where $\langle,\rangle$ is the pairing between cohomology and homology.
In particular, clearly $\langle b(S),Z(S)\rangle=1$ for all
$S\in\Or^k$. 

Combining this observation with the previous proposition, the
statement reduces to showing that there are no $S,Q\in\bcb_\sigma$ and
$\tau\in S_k$ such that 
\begin{equation} \label{flags}
\fl(S^\tau)=\fl(Q).
\end{equation}  Indeed, assume that
such $S,Q,\tau$ exist. First, by the assumption we have
$S^\tau(k)=Q(k)$; denote this element by $H$. Next, again because of
\eqref{flags} the triple $S^\tau(k-1), Q(k-1), H$ must be linearly
dependent. Now if $\sigma(S^\tau(k-1))<\sigma(Q(k-1))$, then this
contradicts $Q\in\bcb_\sigma$, while
$\sigma(S^\tau(k-1))>\sigma(Q(k-1))$ contradicts
$S\in\bcb_\sigma$. Thus the only possibility is
$S^\tau(k-1)=Q(k-1)$. Continuing inductively we can show that
$S^\tau(j)=Q(j)$ for $j=1,\dots,k$, which implies $S=Q$. 
\end{proof}

\begin{cor}
The set $Z(\bcb_\sigma^k)$ is a basis of $H_k(U_\C,\Z)$.
\end{cor}
\begin{rem}
Compare this result with Lemmas 2.10-11 in \cite{falkt}.
\end{rem}

Let us give a name to the property introduced in \propref{id}:
\begin{deff}
A set $DB^k\subset\Or^k$ is a {\em diagonal basis} in degree $k$ if
$$\sum_{S\in DB^k}(b\tensor Z)(S)= \diag^k.$$ 
\end{deff}

\propref{id} shows that any ordering $\sigma$ gives rise to a diagonal
basis $\bcb^k_\sigma$ for $k=1,\dots,n$. In \secref{applic} we will see an
example of a diagonal basis which does not come from a BCB.

Note that \propref{id} also shows, that $D^k_\sigma=D^k_{\sigma'}
\mod \ker(b\tensor Z)$ for any two orderings $\sigma$ and $\sigma'$.
This formally gives us  $|\A|!-1$ relations in 
$\ker(\ir)\subset\Rf^k$ for each $k$.

We make the following
\begin{conj}
The kernel of $\ir$ is generated by the relations $D^k_\sigma\sim
D^k_{\sigma'}$ for all $k$.
\end{conj}

Now, we check how BCBs behave under the standard
deletion-contraction in $\A$. What follows is parallel to the
arguments in \cite{orlik}. 

We take advantage of the usual technique of a triple, basic in the
theory of HPAs \cite[Chapter 3]{orlik}. For $H\in\A$, the corresponding
triple consists of $\A$, $\abez$ and $\A|H$, the arrangement
induced on $H$ by intersections with other elements of $\A$. Thus
there is a natural surjection $\pi:\abez\arr\A|H$. 

Pick an element $H\in\A$ and fix an ordering $\sigma:\A\arr\bara$,
{\em compatible} with $H$ in the following sense:
\begin{enumerate}
\item  $H$ is last, i.e. $\sigma(H)= |\A|$ and
\item if $H_1\cap H = H_3 \cap H$ and $\sigma(H_1) < \sigma(H_2) <
\sigma(H_3) $, then  $H_1\cap H = H_2 \cap H$.
\end{enumerate}
The ordering $\sigma$ induces an obvious ordering $\sigma\backslash H$ on
$\abez$, and in view of the second property also an ordering $\sigma|H$ on
$\A|H$. via the lift Denote by $l_\sigma$ the lift $\A|H\arr\abez$ which
takes an element $H'\in\A|H$ to the element $H''\in\abez$ which satisfies
$\pi(H'')=H'$ and has the {\em least $\sigma$-value}.

\begin{prop} \label{bcbsplit}
$$\bcb_\sigma^k(\A) = \bcb_{\sigma\backslash H}^k(\abez)\cup
l_\sigma(\bcb^{k-1}_{\sigma|H}(\A|H))*H;$$
In words: the BCB of $\A$ is the BCB of $\abez$ plus the set of 
elements of $\Or^k$ obtained by appending $H$ to the end of the lifts
of the BCB of $\A|H$.
\end{prop}
\begin{proof}
The proof is a straightforward check of the definitions based on the
following fact from linear algebra: a set of hyperlanes
$H_1,\dots,H_k,H\in\A$ is linearly independent if and only if the
hyperplanes $H_1\cap H,\dots,H_k\cap H$ in $H$ are linearly independent.
\end{proof}

\section{Integral arrangements and the main result} \label{integral}

Let $\A$ be an integral arrangement. In this section, we describe a
deformation of the constructions of the previous section, which 
uses the extra structure of integrality. Note that most constructions will
work for $k=n$ only.

We retain the notation of \secref{intro} and introduce some more:
 $\ach$ is the dual arrangement in $\vs$,
consisting of the hyperplanes dual to the one-dimensional faces of $\A$;
$\Delta$ will stand for an open chamber in the decomposition of $\vs$ generated
by all lattice translates of these dual hyperplanes, i.e. by
$\Lambda+\cup\ach$. We will call a point $t\in \vs$ in the complement
of $\Lambda+\cup\ach$ {\em regular}, and denote by $\Delta(t)$ the
chamber which contains $t$. 

Our goal is to find a residue form $\omega_\Delta\in
\oma^n\tensor H_n(U_\C,\Z)$ such that for every regular $t\in \Delta$ and
$f\in\rat$ we have
$$B_f(t) = \langle \omega_\Delta,\exp(t)f\rangle.$$

Here we think of $t\in\vs$ as linear function on $V$.

As a first step, we construct a new map $\hos_\Delta:\Or^n\arr\oma^n$ as
follows. Let $S\in \Or^n$, and choose an ordered set of forms
$\vex=(x_1,\dots,x_n)$, so that $x_i\in \Lambda$ and $x_i$ 
defines $S(i)$. Let $\cube(\vex)$ be the unit cube 
$$\{z\in\vs|\,z=\sum \lambda_i x_i,\,0\leq \lambda_i<1\},$$ and denote by
$\vol(\vex)$ the volume of this cube with respect to $\Lambda$. Then define
\begin{equation} \label{cone}
\hos_\Delta(S) = \frac1{\vol(\vex)}\wedge_{i=1}^n
\td(x_i)\sum_{y\in\Lambda\cap(\cube(\vex)-t)}\exp(y), 
\end{equation}
where $t$ is a point in $\Delta$; the form $\td(x)$ was defined in
\lemref{onedim}. Note that the sum in the definition contains $\vol(\vex)$
terms. The definition is justified by the following
\begin{prop} \label{cubeprop} 
The form $\hos_\Delta(S)$ is independent of the 
representative forms $x_i$ and only depends on $S$ and $\Delta$.
\end{prop}
{\bf Example:} Consider the one-dimensional case and let
$t\in(0,1)$. Then we have
$$\frac{dx}{1-\exp(x)}=\frac{1+\exp(x)}{2}\frac{d2x}{1-\exp(2x)}=
\frac{\exp(-x)\,d(-x)}{1-\exp(-x)}.$$ 
To place this statement in the correct framework, it is convenient to
introduce the analogs of the spaces of rational functions and
logarithmic differentials in the integral case.
\begin{deff}
Let $\cla$ be the ring over $\C$ generated by $\{\exp(x)|\,
x\in\Lambda\}$, the coordinate ring of the torus $V/\Gamma$. Let
$\DOS(\A)$ the graded $\cla$-subalgebra of $\oma^*$ generated by the forms $\td(x)$,
for $x\in\Lambda$, defining some hyperplane in $\A$. A related object
is the $\cla$-algebra $\rt$ generated by the functions
$(1-\exp(x))^{-1}$ for the same set of $x\in \Lambda$.
There is an extension of graded algebras
$\hat h:\DOS(\A)\arr\OS(\A)$ taking $\td(x)$ to $d\log(x)$ and
$\exp(x)\in\cla$ to $1\in\C$.
\end{deff}

{\em Proof of \propref{cubeprop}.} For any subset $U\subset\vs$ define
the function $\chi(U)=\sum_{y\in U\cap\Lambda}\exp(y)$ which will be
called the {\em character} of $U$. Also, denote by $d\Lambda_{\vex}$ the volume
form on $\vs$ induced by $\Lambda$ and oriented according to $\vex$.
The proof is based on the following geometric interpretation of \eqref{cone}:
\begin{equation} \label{geometric}
\hos_\Delta(S)=  \chi(-t+\Cone(\vex))\; d\Lambda_{\vex},
\end{equation}
where $\Cone(\vex)$ is the closed cone generated by the $x_i$-s and the
equality means that in the LHS the denominators are expanded according to
$$(1-u)^{-1}=\sum u^i.$$ 

The following is a central technical statement. It shows that the relations
in the algebra $\rt$ reflect the geometry of the cones in the lattice
$\Lambda$. 
\begin{lem} \label{fourier} 
Define $P(\bx)\in\rt$ by 
$$P(\bx)=\prod_{x\in\bx} \frac1{1-\exp(x)}$$ for a linearly independent
subset $\bx\subset\Lambda$.  Let $\bx_\beta, \beta\in \bbeta$ be a finite
collection of linearly independent subsets of $\Lambda$ and
$x_\beta\in\Lambda, \beta\in \bbeta$ such that the disjoint union $\bigcup
(x_\beta+\Cone(\bx_\beta))$ is invariant with respect to some translation
$y\in\Lambda$. Then the following relation holds in $\rt$:
\begin{equation} \label{P}
\sum_{\beta\in \bbeta} \exp(x_\beta) P(\bx_\beta) =0.
\end{equation}
\end{lem}
\begin{proof}
The character $\chi\left(\bigcup (x_\beta+\Cone(\bx_\beta))\right)$ is
exactly the LHS of \eqref{P}. Thus the lemma follows from the fact
that the character of a translation invariant set is a distribution
with support in codimension 1. 
\end{proof}

Now we can complete the proof of the Proposition. The ambiguity in the
definition \eqref{cone} of $\hos_\Delta(S)$ arises because one can multiply
an $x_i$ by an integer. The definition given in \eqref{geometric} is more
invariant, since it does not change if we multiply some $x_i$ by a {\em
positive} integer. It does change if some $x_i$ changes sign, however the
difference between the two formulas will be a volume form with a
coefficient which is the character of an $x_i$-invariant
set. Thus by the Lemma the two definitions coincide. \qed
\medskip

The Lemma can be also used to prove the following deformation of the
circuit relations. Recall the notation of \propref{circuit}. 
\begin{prop}
For any ordered circuit $C$, one has 
\begin{equation} \label{defrelation}
 \sum_{i=1}^{|C|} (-1)^i \mudel(C[i])=0
\end{equation}
\end{prop}
\begin{proof}
 The proof is modeled on the corresponding proof of the circuit
relation (\propref{circuit}) described in the previous paragraph. Indeed,
using \propref{cone} we can again assume that the linear dependence among
the $x_i$-s is of the form $\sum_{i=1}^{n+1} x_i = 0$ by taking if
necessary, suitable integer multiples of the defining forms. Then the
expression in \eqref{defrelation} simply reduces to the sum of characters
of the $n+1$ cones generated by the the $x_i$-s with vertex at $-t$. The
union of these cones is the the whole of $\vs$. Now the proposition follows
from \lemref{fourier}.
\end{proof}

This Proposition is of key importance for us. Combined with \propref{kermuw}
it shows that $\mudel$, just as $\mu$, factors through $b$, thus
there is a map $\qdel: H^n(U_\C,\Z) \longrightarrow \oma^n$ such that
$\mudel = \qdel\circ b$. The following diagram summarizes the
situation. The map $s_\Delta$ is a splitting of the extension $\hat h$.
$$ \begin{diagram}
        &                               &       \oma^n          &\lInto                 &\OS^n          \\
        &\ruTo^\mu                      &       \uInto_q        &\ruTo                  &               \\
\Rf^n   &       \rTo^b                  &        H^n(U_\C,\Z)   &                       &\uTo^{\hat{h}}\dTo_{s_\Delta}  \\
        &\rdTo_{\mudel}                 &       \dInto_{\qdel}  &\rdTo                  &               \\
        &                               &       \oma^n          &\lInto                 &\DOS^n         \\
\end{diagram}
$$

\begin{rem}
Our deformed embedding of the top piece of the Orlik-Solomon algebra into
differential forms seems to be a trigonometric deformation of the
constructions in \S5 of \cite{gelvar}.
\end{rem}
Now we can extend the constructions of the previous section as
follows. Again, tensor this diagram with diagram \eqref{homology}, use the
symbol $\qdel$ for $\qdel\tensor\id$, and introduce the notation $\hatir =
\mudel\tensor Z$.  For every chamber $\Delta$, we obtain special
elements
\[
\qdel(\diag^n)\in \oma^n\tensor H_n(U_\C,\Z),
\]
which can be represented by an ``iterated residue``
\begin{equation} \label{diagir}
\qdel(\diag^n)=\sum_{S\in DB^n}\hatir(S),
\end{equation}
where  $DB^n$ is any diagonal basis.
\begin{thm} \label{thethm}
Let $f\in\rat$ and $t\in\vs$ regular element.
\begin{enumerate}
\item For $f\in\rat$ the distribution $B_f$ defined by the Fourier series
\eqref{bdef} is a polynomial function inside
every chamber $\Delta\in\vs$ of $\Lambda+\cup\ach$. 
\item For $t\in\Delta$ we have the following three formulas for the
 the multiple Bernoulli polynomials:
\begin{equation*}
B_f(t) =   \langle\qdel(\diag^n),\exp(t)f\rangle, \tag{A}
\end{equation*}
\begin{equation*} 
B_f(t) = 
\sum_{S\in DB^n} \langle \hatir(S), \exp(t)f\rangle, \tag{B}
\end{equation*}
\begin{equation*}
B_f(t) = \sum_{S\in DB^n} \ires_S 
\frac1{\vol(\vex)} \prod_{i=1}^n
\todd(x_i)\left(\sum_{y\in\Lambda\cap(\cube(\vex)-t)}\exp(y)\right)\exp(t)f,\tag{C}
\end{equation*}
where $\vex=(x1,\dots,x_n)$ is a sequence of forms representing $S$, $DB^n$
is any diagonal basis in degree $n$ and 
\[
\todd(x) = \frac{x}{1-\exp(x)}
\]
\end{enumerate}
\end{thm}
\begin{proof}
First note that the formulas (A) and (B) are equivalent in view of
\eqref{diagir} and  formula (C) follows from formula (B) by
\propref{simply} and the definition of $\mu_\Delta$ in \eqref{cone}.

We will prove the first part of the theorem and  formula (B) by
induction on the number of hyperplanes $|\A|$.

The starting step of the induction will be the case when $\A$ is {\em
elementary}, i.e. $|\A|=n$.  In this case $\Or^n$ consists of one
element and there is a single iterated residue. For $n=1$ formula (B)
reduces to \lemref{distribution}. For $n>1$, pick some representative
forms $\vex=(x_1,\dots, x_n)$ and denote by $\Lambda_{\vex}$ the
sublattice of $\Lambda$ they generate. Without loss of generality we
can assume that $f= \prod_{k=1}^n x_i^{-\lambda_i}$. If $\Lambda_{\vex} =
\Lambda$ then the infinite sum defining $B_f$ is simply a product of
the one-dimensional sums, thus $B_f(t)= \prod_{k=1}^n
B_{\lambda_i}(t_i)$, where the function $B_k$ was defined in
\eqref{defin} and $t_i$ are the components of $t$ in the basis
${\vex}$. The iterated residues also split into a product of
one-dimensional residues: for example,
for the basic chamber $\Delta=\{x|\, 0<x_i<1\}$ 
\begin{multline}
\langle\hatir(H_1,\dots,H_n),\exp(t)\prod_{k=1}^n x_i^{-\lambda_i} \rangle 
\\ =  \prod_{k=1}^n \res_{H_i} 
\frac{x_i}{1-\exp{x_i}}\exp(\{t_i\}x_i)x_i^{-\lambda_i}.
\end{multline}
Clearly, both formula (B) and part 1 hold.

For the general case $\Lambda_{\vex} \neq\Lambda$, note that if we ignore
the correction factor 
\[
\frac1{\vol(\vex)}\sum_{y\in\Lambda\cap(\cube(\vex)-t)}\exp(y)
\]
in \eqref{cone} for the moment, we obtain the formula for the product
Bernoulli polynomial $B_{f,{\vex}}(t)$, corresponding to the lattice
$\Gamma_{\vex}$, which is dual to $\Lambda_{\vex}$, and, naturally, contains
$\Gamma$.  A simple computation shows that
\begin{equation} \label{finite}
B_f(t) =  
\frac1{\vol(\vex)}\sum_{y\in\Lambda/\Lambda_{\vex}}B_f(t+y).
\end{equation}
This uses the fact that the sum of the characters of a finite group
$G$ (in our case $\Gamma_{\vex}/\Gamma$) is the function on the group
with support at the identity and value $|G|$. It is easy to see then
that \eqref{finite} exactly gives rise to the above correction factor
in the iterated residue formulas.

Now consider an arbitrary $n$ dimensional arrangement $\A$ such that
$|\A|>n$. According to our inductive hypothesis, we can assume that
the theorem holds for all arrangements with fewer hyperplanes than
$|\A|$.

To proceed, we recall an important fact: the partial fraction decomposition
in several dimensions. A weak version of the decomposition theorem is
sufficient for us (cf. \cite[Theorem 5.2]{zelev}).
\begin{lem}
The subset of those elements of $\rat$ which are singular along at most $n$
hyperplanes, span $\rat$.
\end{lem}
According to this Lemma, we can always assume for the purpose of proving
the Theorem that $f\in\rat$ is
singular along at most $n$ hyperplanes.  Since $\A$ is not elementary,
we can pick a hyperplane $H\in\A$, along which $f$ is regular. Let $r_H$
be the restriction map from $\vs$ to $H^*$. Then for $f\in\rat$
regular along $H$ by the definition of $B_f$ we have the key equality:
\begin{equation} \label{reduction}
 B_{f,\A}(t) = B_{f,\abez} - B_{f|H,\A|H}(r_H(t)),
\end{equation}
where $f|H$ is the restriction of $f$ onto $H$.  Note that that the theorem
is assumed to hold for $\A|H$ and $\abez$ via our inductive
hypothesis. Note that we Then
\eqref{reduction} immediately implies Part 1 of the theorem since,
according to the inductive assumption, all the walls which can appear as
singularities of $B_{f,\abez}$ and $B_{f|H,\A|H}(r(t))$, have the form
$\lambda+\check H$ for some $\check H\in \ach$.  To prove formula (B), we
compare \eqref{reduction} with \propref{bcbsplit}.  Choose an ordering 
$\sigma$ {\em compatible} with $H$ (see the definition before
\propref{bcbsplit}). Then 
\begin{multline} \label{soksor}
\sum_{S\in \bcb_\sigma^n(\A)} \langle \hatir(S), \exp(t)f\rangle = \\
\sum_{S\in \bcb_{\sigma\backslash H}^n(\abez)} \langle \hatir(S),
\exp(t)f\rangle  + 
\sum_{S\in \bcb^{n-1}_{\sigma|H}(\A|H)*H} \langle \hatir(S),\exp(t)f\rangle.
\end{multline}
Since the chamber structure of $\abez$ is coarser than that of $\A$, every
chamber $\Delta$ of $\A$ defines a chamber $\Delta\backslash H$ of
$\abez$. Similarly, a chamber $r(\Delta)$ of the arrangement $\A|H$ is
defined.  Then the first term on the RHS of \eqref{soksor} is equal to
\[ B_{f,\abez} = \sum_{S\in \bcb_{\sigma\backslash H}^n(\abez)} \langle 
\widehat{\ir}_{\Delta\backslash  H}(S),\exp(t)f\rangle, \]
by the inductive hypothesis. For the second term note that by the last
point of \remref{restrict} we have
\[ \res_H \todd(x_H) \exp(t) f = - \exp(r_H(t))\, f|H \]
Similarly, taking the correction  factor into consideration one obtains
that the second term on the RHS of \eqref{soksor} is equal to 
\[ - \sum_{S\in \bcb_{\sigma\backslash H}^n(\abez)} \langle
\widehat{\ir}_{r_H(\Delta)}(S),\exp(r_H(t))\,f|H\rangle.
\]
This expression is exactly $B_{f|H,\A|H}(r(t))$ by the inductive
hypothesis. Thus we have shown that formula (B) is compatible with
\eqref{reduction} and this completes the proof.
\end{proof}

\section{Calculations, applications} \label{applic}

\subsection{A more practical formula} Formula (C) in \thmref{thethm} is easy to
implement in actual computations. The best way to do this is to perform a
change of variables, so that each element of the diagonal basis $DB^n$
is transformed into the standard coordinate planes in $\C^n$.

Let $\bz = (z_1,\dots,z_n)$ be the sequence of standard coordinates on
$\C^n$, and let $g$ be a meromorphic function. Then as in \secref{itres} define
$$  \ires_{\bz} g = \res_{z_n=0}\dots\res_{z_1=0} g.$$
Introduce
$$ \on{Todd}(\bz) = \prod_{i=1}^n \frac{z_i}{1-\exp(z_i)}.$$

If we perform the substitution $\{x_i=z_i, \,i=1,\dots,n\}$ in each term of
formula (C), then the function $f$  transforms into some function
$f_{\vex}$ on $\C^n$, $t$ becomes a vector $t_{\vex}$ and the set
$(\cube(\vex)-t)\cap\Lambda$ transforms into a set $\lambda_{\vex}$ of
$\vol(\vex)$ 
vectors with rational coefficients.  Then (C) can be rewritten as follows:
\begin{equation} \label{substit}
B_f(t) = \sum_{S\in DB^n}\frac1{\vol(\vex)}\ires_{\bz} \on{Todd}(\bz)
\left[
f_{\vex}(z)\exp(t_{\vex}\cdot z)\sum_{y\in \lambda_{\vex}} \exp(y\cdot z)
\right] ,
\end{equation}
where the $a\cdot b$ stands for scalar product in $\C^n$.
As $S$ runs through  $DB^n$, the part $\ires_{\bz}\on{Todd}(\bz)$ remains
unchanged in this formula, while the part in square brackets changes according to a set of linear transformations.

\subsection{Example: the Lie arrangements} \label{Lie}
We mentioned in the introduction, that our interest in this problem
stems from an attempt to understand the formulas of Witten for the
volumes of parabolic moduli spaces of principal $G$-bundles over
Riemann surfaces. Denote the genus of the surface by $g$ and consider
the case of one puncture. Then the relevant parameters of the problem
are $g$, the genus, and a conjugacy class $\ttt\in G/\on{Ad}G$.

The volume formulas can be described as follows. Fix a Cartan
subalgebra of the Lie algebra of $G$, denote the set of roots by
$\delta$, the set of positive roots by $\delta^+$ and let $(,)$ be the
basic invariant bilinear form. Let $\A_G$ be the integral arrangement
with $V$ the dual of the Cartan subalgebra, $\Gamma$ the weight
lattice; the arrangement is the set of hyperplanes perpendicular to the root
vectors in $V$. Consider the $\A_G$-rational function
$$W(\gamma)=\prod_{\alpha\in\Delta^+} (\alpha,\gamma)^{-1}.$$ Up to a
$\rho$-shift and a constant this function gives the dimension of the
irreducible representation of $G$ of highest weight $\gamma$, when
$\gamma$ is a dominant weight. Note that $W$ is Weyl-anti
symmetric. According to \cite{witten, kef} the volume of the
$[g,\ttt]$-moduli space (up to some normalization) is given by
$$\on{Vol}(g,\ttt) = B_{W^{2g-1}}(t),$$
where $t$ is a dominant representative of $\ttt$.
A detailed study of this formula will be given elsewhere. 

\subsection{The  \sun arrangement} Again let $\{z_i\}$ be the standard
coordinates on $\C^{n+1}$. The arrangement $\dn$ is defined as follows: $V$
is the hyperplane $\sum z_i =0$ in $\C^{n+1}$, $\Gamma=V\cap\Z^{n+1}$ and the
hyperplanes $H_{ij}$ are given by the forms
$x_{ij}=z_i-z_j$. This arrangement is the Lie
arrangement corresponding to the group \sun; in particular, it has a
Weyl-symmetry with respect to the symmetric group $S_{n+1}$.

There are several natural orderings on this arrangement, but there is
a special diagonal basis $\bbb\subset\Or^n$ with some beautiful
properties which is not a BCB. 
To every $\tau\in S_n$ we can associate
an element $B^\tau\in\Or^n$ given by
$$ B^\tau = (H_{\tau(1)\tau(2)},\dots,
H_{\tau(n-1)\tau(n)},H_{\tau(n)(n+1)}).$$ 
Then $\bbb=\{B^\tau|\,\tau\in S_{n}\}$. It is easy to check that
$\bbb$ is indeed a diagonal basis. It comes with a canonically
associated set of representative forms 
$${\vex}^\tau=(x_{\tau(1)\tau(2)},\dots,x_{\tau(n-1)\tau(n)},
x_{\tau(n)(n+1)}).$$ The existence of such a basis was pointed out in
\cite{zelev} and goes back to a construction of Lidskii \cite{lid}.  

The most important property of $\bbb$ is that the group $S_n$,
embedded into the Weyl group $S_{n+1}$, acts on it transitively. Since
this action also preserves the lattice, it is easy to see that the
substitutions in \eqref{substit} are simply implemented by the natural
action of  $S_{n}$ on the expression in square brackets. Also, since
$\vol({\vex}^\tau) = 1$, the sum in the square brackets will consist of
only one term. If we assume that $f$ is Weyl-symmetric and that
the series \eqref{bdef} defining $B_f(t)$ is absolutely
convergent, then we obtain particularly simple formulas. For example, we
have 
\[
B_f(0) = n!\langle\hatir(B^e),f\rangle.
\]
where $B^e$ is the element of $\Or^n$ corresponding to the trivial
permutation, and $\Delta$ is any chamber which has $0$ as a vertex.
With a natural choice of a chamber the formula expands into
\begin{multline} \label{sun}
B_f(0) = n! \ires_{{\vex}} \on{Todd}({\vex}) f = \\
n!\res_{x_n}\dots\res_{x_1} \left(\prod_{k=1}^n
 \frac{x_k}{1-\exp(x_k)}\right) f(x_1,\dots,x_n).
\end{multline}
If we insert the $t$-dependence, the formula becomes more
complicated:
\begin{equation} \label{general}
 B_f(t) = \langle\hatir(B^e),f\sum_{\tau\in S_{n-1}}\exp(\{t^\tau\})\rangle,
\end{equation}
where $\{t^\tau\}$ denotes the $\Lambda$-translate of $t^\tau$ which lies
in the unit cube, defined by the $n$-tuple ${\vex}^e$. For special values of
$t$ this formula simplifies. It reduces to
\begin{equation} \label{central}
 B_f(c) = n!\langle\hatir(B^e),f\exp(c)\rangle,
\end{equation}
for $c=\gamma(1,1,\dots,1,-n)$, for $0<\gamma<1/n$, and for 
$$ c = \left(\frac k{n+1},\dots,\frac
k{n+1},\frac{k-n-1}{n+1},\dots,\frac{k-n-1}{n+1}\right), \quad 1\leq k\leq
n.$$ Formula \eqref{sun} was first published in \cite{me} in a slightly
different form. It was later used and extended by other authors
\cite{jef,mori}, and we have incorporated some of their improvements.

\end{document}